\documentclass[12pt]{article}

\usepackage{times}
\usepackage{amsmath}
\usepackage{graphicx}

\newenvironment{sciabstract}{%
\begin{quote} \bf}
{\end{quote}}

\usepackage[font=small]{caption}

\newcommand{\Om}{\Omega}
\newcommand{\suppmat}{Supplementary Information}

\newcounter{lastnote}

\title{Learning to represent signals spike by spike}

\author
{Wieland Brendel,$^{1,2,3,\ast}$ Ralph Bourdoukan,$^{2,\ast}$ Pietro
  Vertechi,$^{1,2,\ast}$\\ Christian K. Machens,$^{1,\dagger}$,
  Sophie Den\`eve$^{2,\dagger}$
\
\\
\\
\small{$^1$Champalimaud Neuroscience Programme, Champalimaud
  Foundation, Lisbon, Portugal}\\
\small{$^2$Group for Neural Theory, INSERM U960, D\'epartement d'Etudes
  Cognitives,}\\
\small{Ecole Normale Sup\'erieure, Paris, France}\\
\small{$^3$Werner Reichardt Centre for Integrative
  Neuroscience,}\\
\small{University of T\"ubingen, Germany}\\
\small{$^\ast$These authors contributed equally}\\
\small{$^\dagger$To whom correspondence should be addressed;}\\
\small{E-mail: sophie.deneve@ens.fr or christian.machens@neuro.fchampalimaud.org}\\
}

\date{}

\begin{document}

\baselineskip16pt

\maketitle

\begin{sciabstract}
A key question in neuroscience is at which level functional meaning
emerges from biophysical phenomena. In most vertebrate systems,
precise functions are assigned at the level of neural populations,
while single-neurons are deemed unreliable and redundant.
Here we challenge this view and show that many
single-neuron quantities, including voltages, firing thresholds, excitation,
inhibition, and spikes, acquire precise functional meaning
whenever a network learns to transmit information parsimoniously and
precisely to the next layer. Based on the hypothesis that neural
circuits generate precise population codes under severe
constraints on metabolic costs, we derive synaptic plasticity rules
that allow a network to represent its time-varying inputs with maximal
accuracy. We provide exact solutions to the learnt optimal states, and
we predict the properties of an entire network from its input
distribution and the cost of activity. Single-neuron variability and
tuning curves as typically  observed in cortex emerge over the course
of learning, but paradoxically coincide with a precise, non-redundant
spike-based population code. Our work suggests that neural circuits operate far
more accurately than previously thought, and that no spike is fired in vain.
\end{sciabstract}

\paragraph*{}

Many neural systems encode information by distributing it across the
activities of large populations of spiking neurons. A lot of work has
provided pivotal insights into the nature of the resulting population codes \cite{georgopoulos1986neuronal,simoncelli2001natural,schneidman2006,averbeck2006neural,wohrer2013population}
and their generation through the internal dynamics of neural networks
\cite{amari1977dynamics,sompolinsky1995,eliasmith2005,burak2012}.
However, we understand surprisingly
little about the precise role of each individual spike in distributing
information and in mediating learning.

We revisit this problem by studying a population of excitatory (E)
neurons that are interconnected with inhibitory (I) interneurons (Fig.~1Ai).
The excitatory neurons receive many input signals from other neurons
within the brain. To encode these signals efficiently,
each spike fired by an excitatory neuron should ideally contribute new
and unique information to the population code.
If each neuron receives a different input signal, this is easy. However,
if two excitatory neurons receive similar inputs, they need to communicate
with each other so as to not fire spikes for the same type of
information. One possibility is that the inhibitory interneurons
arbitrate such conflicts by creating competitive interactions
between excitatory neurons \cite{boerlin2013predictive}.
How can neurons learn this from experience?

To formalize the problem, we will define a measure for the
coding efficiency of a neural population (see
\suppmat\ for mathematical details).
First, we impose that any downstream area should be able
to decode the input signals, $x_j(t)$, from a weighted sum of the
neural responses, $\hat{x}_j(t)=\sum_{k=1}^N D_{jk}r_k(t)$,
where $D_{jk}$ is a decoding weight, and $r_k(t)$ is the
postsynaptically filtered spike train of the $k$-th excitatory neuron.
Second,
we assume  that the neurons fire as few spikes as possible, or, more generally,
that they minimize a cost associated with firing, which we denote by
$C(r)$. In other words, we measure the efficiency of the population
code through an objective function that trades off accuracy for cost;
this objective function is simply the sum of the coding error and
the cost, $E = \sum_j (x_j - \hat{x}_j)^2 + C(r)$.

The key problem is that a single excitatory neuron has no access to
this global objective function.  Rather, it has access to the input
signals that arrive via feedforward synapses, $F_{ij}$, and to the
filtered spike trains of other neurons that arrive via recurrent synapses,
$\Omega_{ik}$. However, imagine that we could set these
recurrent synapses to be the feedforward weights multiplied by
the decoding weights, so that $\Omega_{ik} = \sum_j F_{ij}D_{jk}$.
If our neurons are leaky integrate-and-fire neurons, and if we
treat the inhibitory interneurons as simple relays for now
(Fig.~1Aii,iii), then the membrane potential of each neuron becomes
$V_i(t) = \sum_j F_{ij} \big(x_j(t)-\hat{x}_j(t)\big)$.
Accordingly, the membrane potential now reflects a part of the
global coding error, {\em despite} being computed from only
feedforward and recurrent inputs. Each time this error becomes too large, the membrane potential
reaches threshold. The neuron fires, updates the decoded input
signal, and thereby decreases the error, as reflected in the voltage
reset after a spike. Furthermore, through the recurrent synapses,
the neuron will communicate the change in the global coding
error to all neurons with similar feedforward inputs. In turn,
any excitatory feedforward input into a neuron will immediately
be counterbalanced by a recurrent inhibitory input (and vice versa).
This latter reasoning links the precision of each neuron's code
to the known condition of excitatory and inhibitory balance
(EI balance) \cite{deneve2016efficient,van1996chaos,amit1997model,shadlen1998,renart2010}. Indeed, balancing
excitatory and inhibitory inputs optimally would minimize the
variance of the membrane potential, and thus, the error projected
in the direction of each neuron's feedforward weights. In other words, EI balance ensures that   
$V_i(t) = \sum_j F_{ij} \big(x_j(t)-\hat{x}_j(t)\big) \approx 0$
\cite{boerlin2013predictive,boerlin2011}.

\begin{figure}
\centering
\includegraphics[width=16cm]{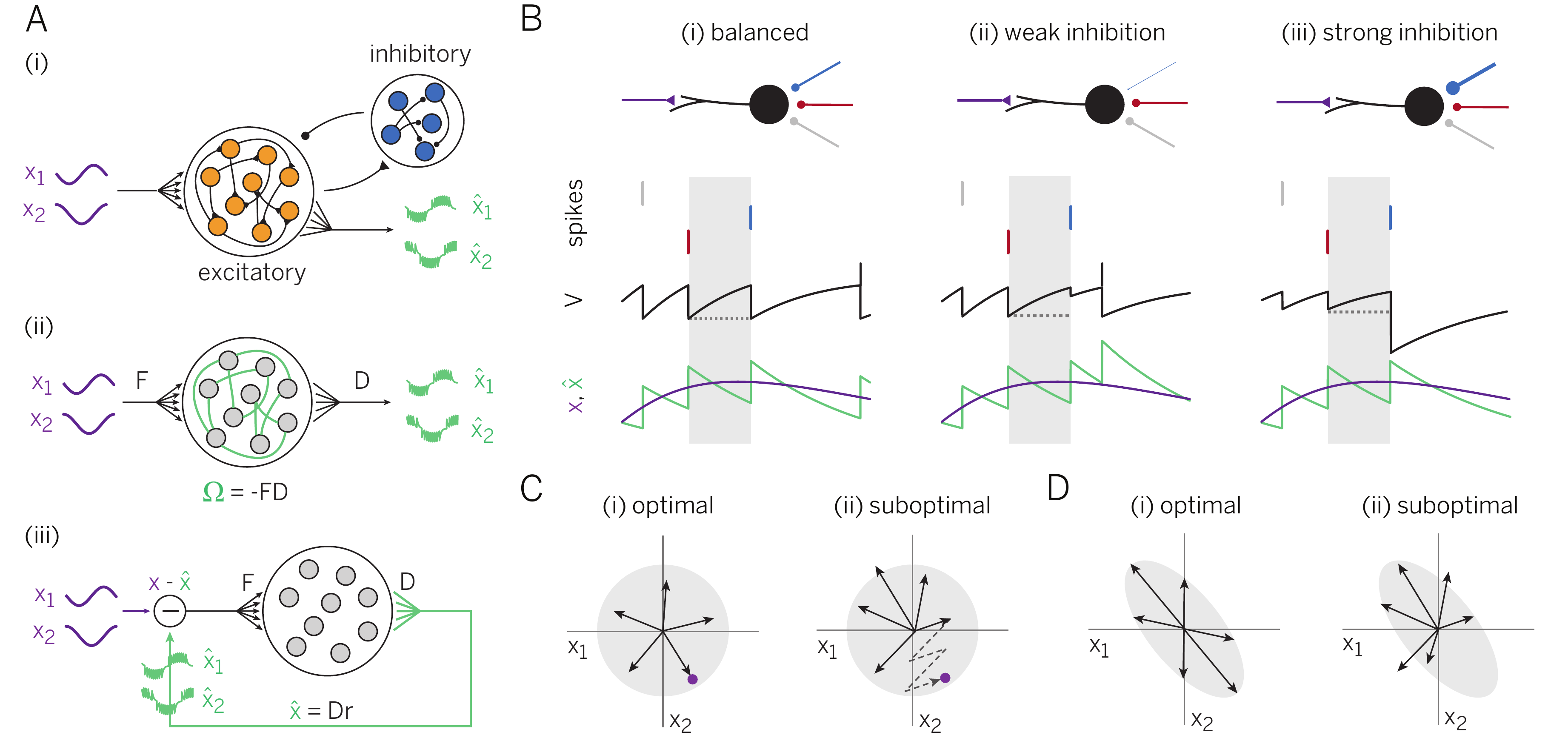}
\caption{Networks learning to represent analog signals efficiently with spikes.
{\bf A.}
(i) Recurrent neural network with input signal $x$ (purple) and
signal estimate $\hat{x}$ (green), as read out from the spike trains of the
excitatory population.
(ii) Simplified network without separate excitatory and inhibitory populations.
($F$=feedforward weights, $D$=decoding weights, $\Omega$=recurrent weights)
(iii) Same as (ii), but unfolded to illustrate the effect
of the recurrent connections.
{\bf B.} A single neuron's EI balance as a target of learning for recurrent connections.
(i) Ideal case with EI balance.
(ii) One inhibitory synapse too weak.
(iii) One inhibitory synapse too strong.
Shown are the neuron's membrane voltage (black), spikes
from three inhibitory neurons (vertical lines, color-coded
by connection), signal (purple), and signal estimate (green).
{\bf C.} Influence of feedforward weights on signal encoding
and decoding, shown for a five-neuron network encoding
two signals with zero mean and equal variance (gray area).
(i) Optimal scenario.
(ii) Sub-optimal scenario.
{\bf D.} Similar to {\bf C}, but for correlated input signals.}
\end{figure}

How can a network of neurons learn to move into this very specific
regime? Several learning rules for EI balance have been successfully
proposed before \cite{song2000competitive,vogels2011inhibitory},
and spike-timing-dependent plasticity (STDP) can even balance EI currents
on a short time scale \cite{vogels2011inhibitory}. However, here we
both need to balance EI currents as precisely as possible, and we need
to ensure convergence onto the right type of recurrent connectivity
(Fig.~1Aii,iii).  We will first examine the problem of EI balance
more carefully by studying a neuron's membrane potential directly
after it receives an inhibitory spike from one of its recurrent connections
(Fig. 1Bi; black trace). After the inhibitory spike (Fig. 1Bi, red),
the neuron integrates its (excitatory) feedforward input currents, leading to a transfer of
electric charges across the membrane. The arrival of the second
(inhibitory) spike
(Fig. 1Bi, blue) then causes a transfer of charges in the opposite
direction, which ideally should cancel the total charge accumulated through the
feedforward inputs. When the inhibitory spike overshoots (undershoots)
its target, then the respective synapse needs to be weakened
(strenghtened), see Fig 1Bii,iii. This scheme keeps the neuron's
voltage (and thereby the coding error) perfectly in check. The regime
can be reached by a simple voltage-based learning rule for the
recurrent weights of neuron $i$, applied each time a presynaptic
neuron $k$ spikes,
\begin{equation}
\Delta\Om_{ik} \propto -\beta(V_i + \mu r_i) - \Om_{ik} - \mu\delta_{ik}.
\end{equation}
Here $V_i$ is the postsynaptic membrane potential before the
arrival of the presynaptic spike, while $\beta$ and $\mu$ are positive terms
that implement a possible cost factor $C(r)$ (see \suppmat).


Fig.~2 illustrates the effect of this learning rule in a network
with 20 neurons receiving two random, time-varying inputs. Here
the network was initialized with lopsided feed-forward weights
and with recurrent weights equal to zero (Fig.~2Bi). While
the network receives the random inputs, the recurrent
synapses change according to the learning rule, Eq.~1, and
each neuron thereby learns to balance its inputs. Once learnt,
the recurrent connectivity reaches the desired structure,
$\Om_{ik} = -\sum_j F_{ij}D_{jk}$ for some $D_{jk}$,
and the voltages of the neurons become proportional to
part of the coding error (see \suppmat\ for convergence proof).
As a result of the EI balance, the voltages fluctuations of
individual neurons are much better bounded around the resting
potential (Fig.~2Eii), the global coding error decreases
(Fig.~2Aii,Cii), and the network experiences a large drop in the
overall firing rates (Fig.~2Aii,Dii).

Despite these overall improvements, however, the network still
fails to represent part of the input, even after the recurrent
connections have been learnt (Fig.~2Cii, arrow). Indeed, in the
example provided, this part of the input signal cannot
be properly represented because the feedforward connections
do not cover the full two-dimensional signal space (Fig.~2Bii),
which becomes particularly evident in the tuning curves of the
individual neurons (Fig.~2Gii).

Consequently, the feedforward connections need to change as well,
so that all parts of the input space are properly covered. We can
again obtain a crucial insight by considering the
final, `learnt' state, in which case the feedforward connections
are directly related to the optimal decoding weights. For example, for uncorrelated inputs, 
the optimal feed-forward and decoding weights are equal, i.e. 
$F_{ik}=D_{ki}$ (see \suppmat).
In Fig.~1C, we examine the decoding problem from the point of
view of five neurons that seek to represent two input signals.
If an input signal lies approximately in the direction of the
vector of one of the neurons' decoding weights, then a few spikes suffice to
represent it accurately (Fig.~1Ci). If the input signal lies elsewhere,
many more spikes are required to achieve the same
accuracy (Fig.~1Cii). For random input signals with zero mean and equal variance,
as in Fig.~1Ci, the best representation is achieved
when the decoding vectors are evenly distributed. 
(see \suppmat\ for details and convergence proofs). For correlated inputs, 
the decoding weights should provide optimal coverage by favoring more frequent input signal directions 
(See Fig.~1Cii and \suppmat).   

The feedforward weights of neuron $j$ can learn to optimally cover
the input space if they change each time neuron $j$ fires a spike,
\begin{equation}
\Delta F_{ij} \propto \alpha x_i - F_{ij},
\end{equation}
where $x_i$ is the feed-forward input signal, and $\alpha>0$ is a scaling
factor. In the case of correlated inputs, the term ``$F_{ij}$'' is replaced by the co-variance of pre and post-synaptic input currents. 

From the perspective of standard frequency-modulated plasticity, the
learning rule is Hebbian in that connections are reinforced for
co-occurring high pre- and postsynaptic activity. Because of the
competition introduced by the recurrent connections, a post-synaptic
spike occurs only if no other neuron fired first in response to the
same signal. This introduces repulsion between the feedforward
weights of different neurons and eventually leads to an even
coverage of the input space.

\begin{figure}
	\centering
	\includegraphics[width=13.5cm]{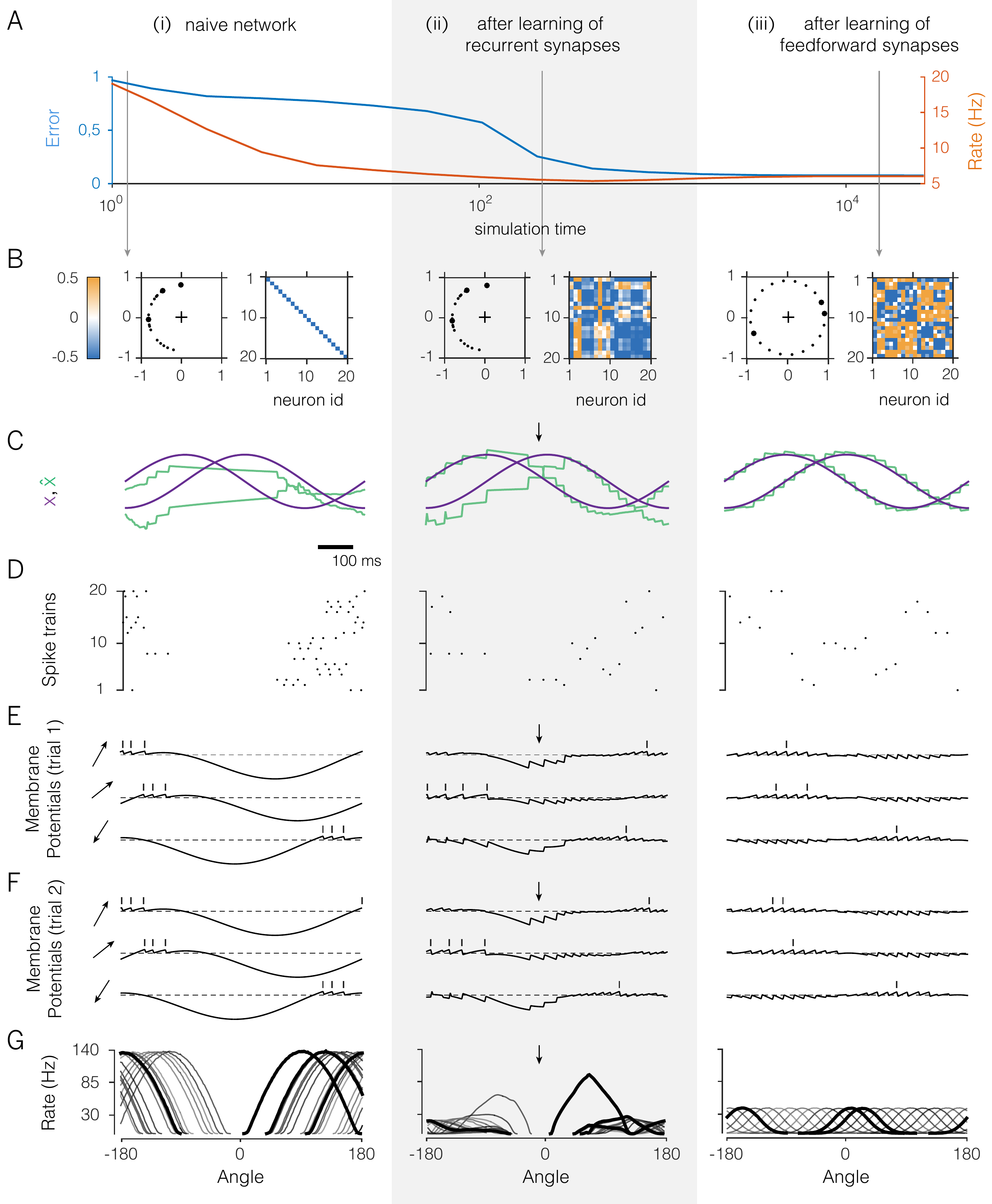}

\caption{A 20-neuron network that learns to encode two signals.
{\bf A.} Evolution of coding error (blue) and mean population firing rate
(orange) over learning.
{\bf B.} Feedforward and recurrent connectivity at three stages
of learning. In each column, the left panel shows the two-dimensional
feedforward weights, and the right panel the matrix of
recurrent weights. Diagonal elements correspond to the neurons'
self-resets after a spike.
{\bf C.} Example of time-varying input signals (purple) and
signal estimates (green). Signal estimates in the naive network are
constructed using an optimal linear decoder.
Arrows indicate parts of the signal space that remain poorly
represented, even after learning of the recurrent weights.
{\bf D.} Spike rasters from the network.
{\bf E.} Voltages and spike times of three exemplary neurons. Dashed
lines illustrate the resting potential.
{\bf F.} As in {\bf E}, but for a different trial.
{\bf G.} Tuning curves (firing rates as a function of an input
with variable angle and constant radius in polar coordinates)
of all neurons in the network.}
\end{figure}

The effect of the feedforward plasticity rule is shown in Fig.~2Aiii--Giii.
The feedforward weights change until the input space is spanned
more uniformly (Fig.~2Biii). While these changes are occuring, the
recurrent weights remain plastic and keep the system
in a balanced state. At the end of learning, the neuron's tuning curves
are uniformally distributed (Fig.~2Giii), and the quality of the
representation becomes optimal for all input signals (Fig.~2Aiii,Ciii).

Importantly, the final population code represents the input signals
spike by spike, with a precision that approaches the discretization
limit imposed by the spikes. Initially, the neurons are unconnected (Fig. 2Bi),
and their voltages reflect the smooth, time-varying input
(Fig. 2Ei,Fi). Moreover, neurons fire the spikes at roughly the same
time from trial to trial. After learning,
the membrane potentials are correlated, reflecting
their shared inputs, yet the individual spikes are far more susceptible
to random fluctuations (Fig. 2Eiii,Fiii). Indeed,
whichever neuron happens to fire first immediately inhibits (resets)
the others, so that a small initial difference in the membrane
potentials is sufficient to change the firing order completely.
The random nature of spike timing is therefore a direct
consequence of a mechanism that prevents any
redundant (or synchronous) firing. More generally, any source of noise or
dependency on previous spike history will change the firing order, but
without a significant impact on the precision of the code. Thus, variable spike trains
co-exist with a highly reproducible and precise population code.

Fortunately, the same results can be obtained in networks with
separate excitatory (E)  and inhibitory (I) populations (Fig.~1Ai).
In this more realistic case,  the inhibitory population must simply
learn to represent the population response of the excitatory
population, after which it can balance the excitatory population in turn.
This can be achieved if we train the EI connections using the
feedforward rule (Eq.~2) while the II, EE, and IE connections
are trained using the recurrent rule (Eq.~1; see \suppmat\ for
details).

\begin{figure}
\centering
\includegraphics[width=15cm]{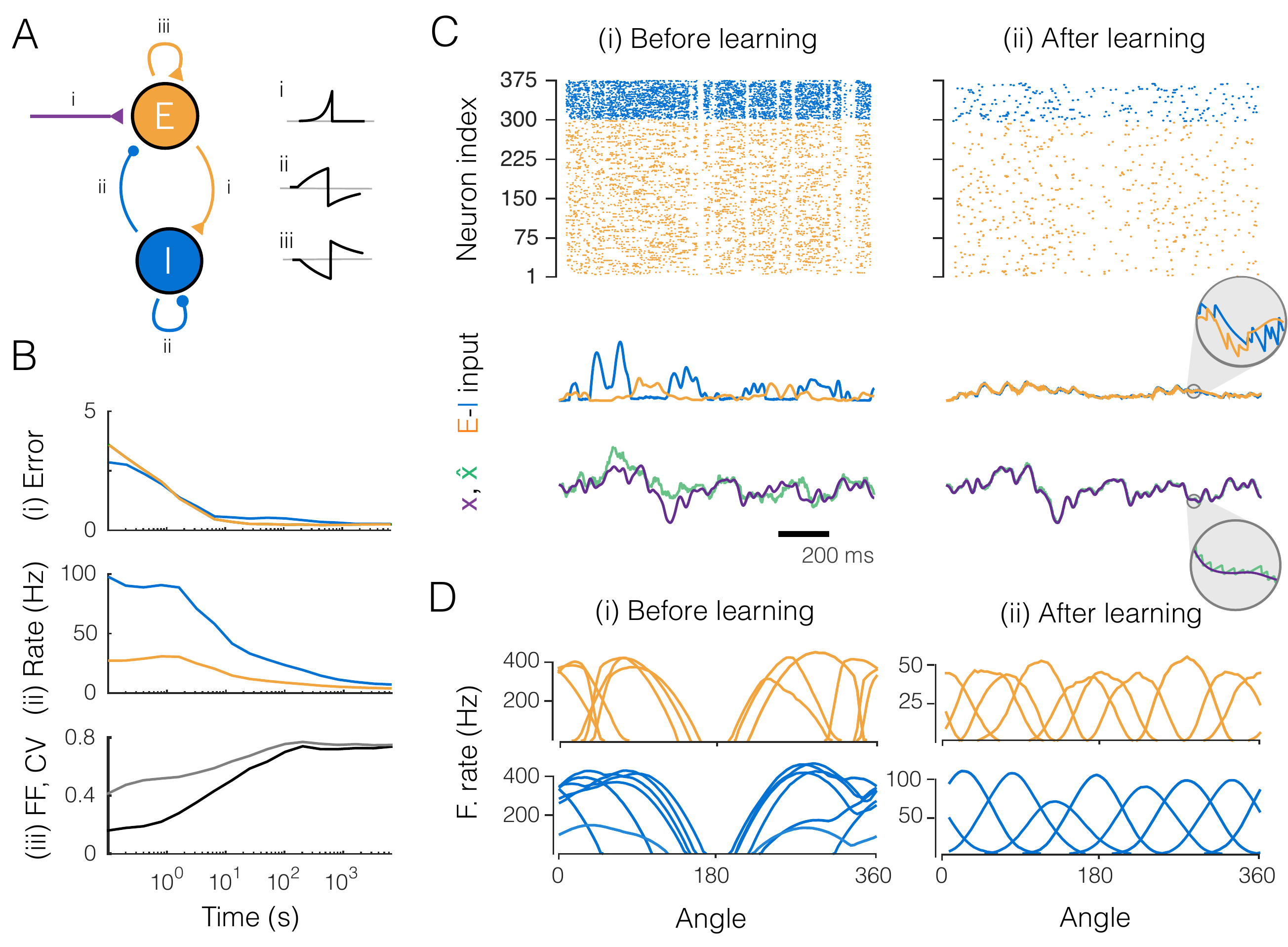}

\caption{Large network (300 excitatory and 75 inhibitory neurons)
that learns to encode three input signals. Excitation shown
in orange, inhibition in blue.
{\bf A.} The EI network as in Fig.~1Ai and the learning rules (i,
feedforward rule; ii and iii, recurrent rule). The insets
show the results of an STDP-like protocol between pairs of
neurons applied to our learning rules, with the x-axis
representing the relative timing between pre- and post-synaptic spikes,
and the y-axis the change in weight.
{\bf B.}
Evolution of the network during learning.
(i) Coding error for excitatory and inhibitory populations. Black line shows the decoding error 
for a population of 400 independent Poisson firing neurons with identical instantaneous firing rates.
(ii) Mean firing rate of excitatory and inhibitory populations.
(iii) Averaged coefficient of variation (CV, gray) and Fano factor
(FF, black) of the spike trains.
{\bf C.} Network input and output before (i) and after (ii) learning.
(Top) Raster plots of spike trains from excitatory and
inhibitory populations.
(Center) Excitatory and inhibitory currents into one
example neuron. After learning, inhibitory currents tightly balance
excitatory currents (inset).
(Bottom) One of the three input signals (purple) and the corresponding
signal estimate (green) from the excitatory population.
{\bf D.} Tuning curves (firing rates as a function of the angle for
two of the input signals, with the third signal clamped to zero)
of the most active excitatory and inhibitory neurons.}
\end{figure}

Fig.~3 illustrates how the key results obtained in Fig.~2 hold in the the full EI network.
The network converges to the optimal balanced state (Fig.~3B),
and the precision of the representation improves substantially
and approaches the discretization limit (Fig.~3Bi, Cii), despite the overall
decrease in output firing rates (Fig.~3Bii, Cii). Initially
regular and reproducible spike trains (Fig.~3Biii) become asynchronous, irregular, and comparable
to independent Poisson processes (Fig.~3Biii, pairwise
correlations are smaller than 0.001). Furthermore, we observe that the
neuron's tuning curves, when measured along the first two signal directions,
are bell-shaped just as in the previous example
(Fig.~3Dii). Note that the inhibitory neurons fire more and have
broader tuning than the excitatory neurons. This result holds
independent of the chosen initial state of the network, and is
simply owed to their smaller number.

Finally, we tested whether the network was robust to other limitations
of biological microcircuits, including strong noise injected in the
membrane potential, realistic synaptic dynamics with transmission
delays, or even randomly removing half of the connections between
neurons.  We found that under a wide range of conditions, the
network learnt to achieve a performance near the discretization limit,
outperforming conventional spiking networks or rate-based population coding models (see supplementary
Fig.~S1). Our findings suggest that, despite being derived for an
all-to-all connected network with instantaneous synapses, the learning
rules can tolerate large deviations from such initial conditions. This
robustness is inherited from the generality of the relationship
between E/I balance and efficiency and the subsequent error-correcting
coding strategy in the network \cite{barrett2016optimal}.

We have so far considered uncorrelated inputs. For correlated input
signals, the network learns to align its decoding weights to the more
frequent signal directions (Fig.~4A). As a result, the tuning curves of the
learnt network reflect the distribution of inputs experienced by the network
(Fig.~4B). In particular, tuning curves are denser and sharper for signal directions that are a-priori more probable. 
This result is reminiscent of the predictions for efficient rate-based population codes with independent Poisson noise \cite{Ganguli2014}. Note, however, 
that our networks learn a spike-per-spike code far more precise and efficient than such rate-based population codes. 

\begin{figure}
\centering
\includegraphics[width=15cm]{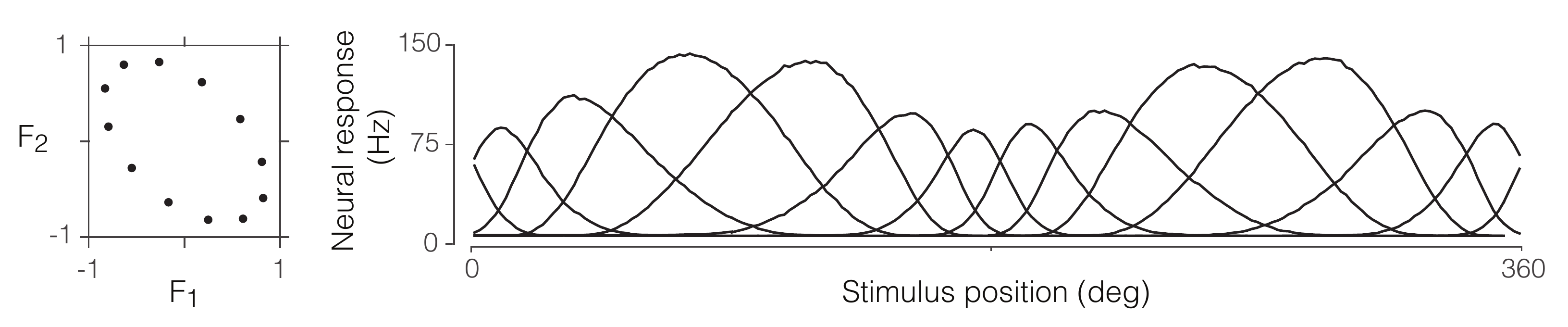}

\caption{Example of a 12-neuron network that learns to encode two correlated input signals (with a distribution similar to Fig. 1D).
{\bf A.} The two-dimensional feedforward weights of the 12 neurons after learning. 
{\bf B.} Tuning curves of all neurons in the network after training, i.e., their firing rates as a function of the angle of a two-dimensional input with constant radius in polar coordinates.}
\end{figure}

To further demonstrate the power of the learning
rules, we trained a network to represent speech signals, filtered
through 25 frequency channels, in its spiking output (Fig.~5A).
Despite consisting of 100 neurons that fire at only $\sim 4$Hz
the network learns to represent the signals with high precision
(Fig.~5B,C). This feat would be impossible if the network had
not learnt the strong correlations in speech. 
As a drawback,  the network has become specialized, and a new
``non-speech'' stimulus results in poor EI balance, high firing
rates, and poor coding (Fig.~5D,E). After experiencing the new
sound several times, however, the network represents the
``non-speech'' sound as precisely and parsimoniously as the
previously experienced speech sounds (Fig.~5F).

After training with the speech signals, the feedforward and decoding weights 
adopt a structure reflecting the natural statistics of speech. 
The feedforward weights typically have excitatory subfields covering
a limited range of frequencies, as well as inhibitory subfields (Fig.~6A). Decoding weights are wider and more complex, thus exploiting the high 
correlations between frequency channels (Fig.~6B). These model predictions are broadly compatible with observations in the mammalian auditory pathway, and notably the representation of speech signals in A1 \cite{Mesgarani2014}.  After retraining to the new stimulus, feed-forward weights are modified specifically at the frequencies of the new stimulus (Fig.~6A). However, these changes are not massive. In particular, only a handful of neurons (two in this example) have become truly specialized to the new stimulus, as reflected by their decoding weights ( Fig.~6B). 

The accommodation of the network to the new stimulus is largely
mediated by plasticity at the recurrent synapses, whereas the feedforward
synapses are less essential.
Indeed, turning off feedforward plasticity (but not recurrent plasticity)
lets the network reach almost the same performance for the new stimulus,
whereas turning off recurrent plasticity (but not feedforward
plasticity) can even worsen the coding performance instead
of improving it (Fig.~6C).

\begin{figure}
\centering
\includegraphics[width=11cm]{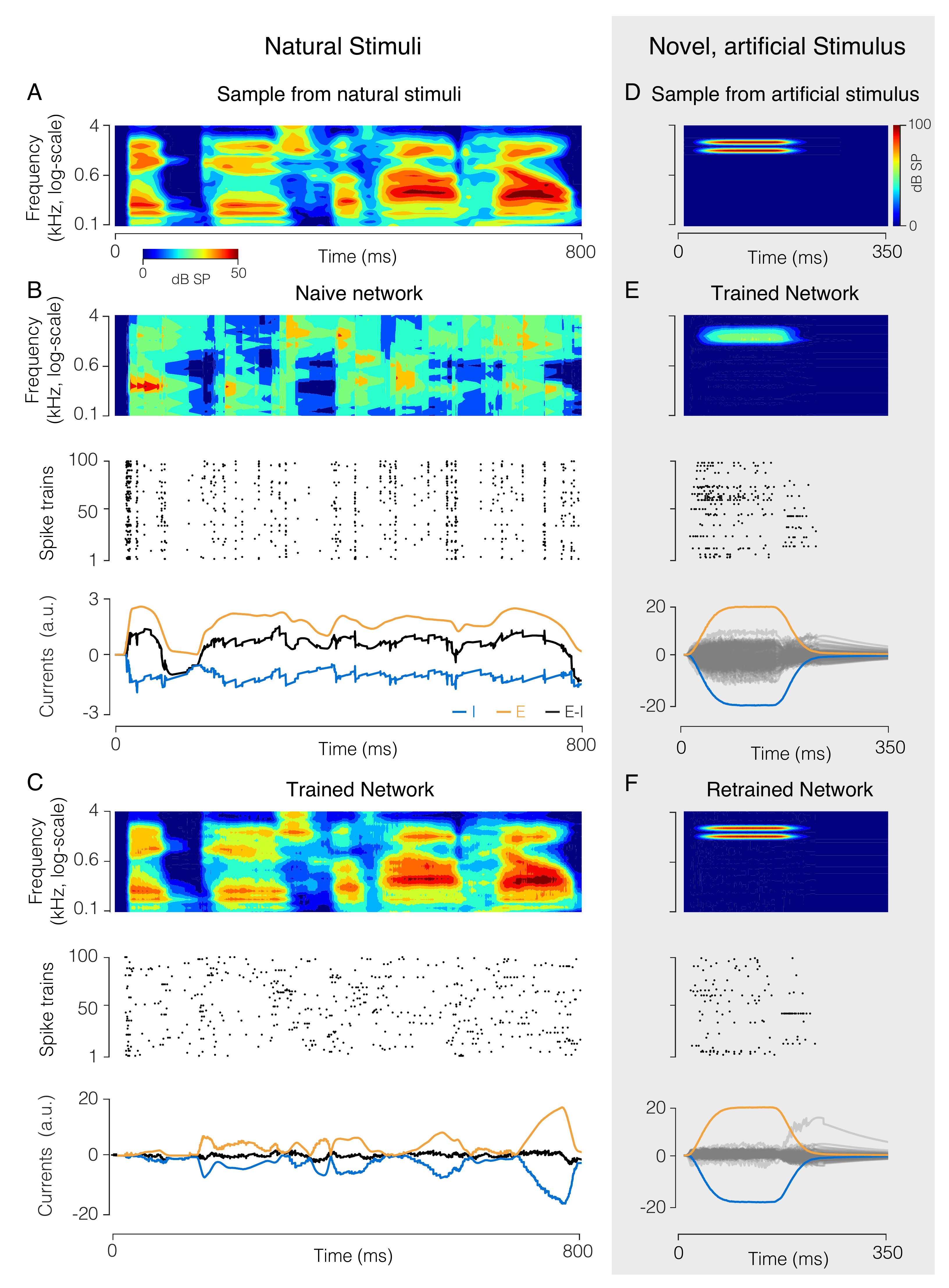}

\caption{Network (100 neurons) that encodes a high-dimensional, structured
natural input (speech sounds).
{\bf A.} Spectrogram of a speech sound.
{\bf B.} ``Naive'' network with random feedforward and
recurrent weights.
(Top) Optimal linear estimator applied to output spike trains
reconstructs the stimulus poorly.
(Center) Spike raster from all neurons, showing synchronous firing.
(Bottom panel) Excitatory (orange) and inhibitory (blue) current
into an example neuron are poorly balanced, causing large fluctuations
in the total current (black).
{\bf C.} Same as {\bf B}, after learning. The signal estimate
tracks the signal closely (top), spike trains are asynchronous and
irregular (center), and EI currents are tightly balanced (bottom).
{\bf D.} Spectrogram of artificial, ``non-speech'' sound.
{\bf E.} Response of the trained network trained to a non-speech
sound, similar format as {\bf B}, {\bf C}. The new sound is improperly
reconstructed (top), and EI responses are poorly balanced (bottom).
Grey lines show the superposed total currents for all neurons,
orange and blue lines show the mean excitatory and inhibitory
currents, averaged over the population.
{\bf F.} Same as {\bf E}, after re-training the network with a mixture
of speech sounds and the new sound. The new sound is now
represented precisely (top) with fewer spikes (center), and EI
balance is improved (bottom).}
\end{figure}

Since learning the new stimulus relies on the recurrent connections
re-balancing the feed-forward inputs, EI balance should directly reflect
behavioral performance, a prediction compatible
with recent observations in the auditory cortex \cite{Marlin2015}. As a
consequence, blocking inhibitory plasticity during
perceptual learning should result in a worsening of EI balance
and behavioral performance, while blocking excitatory plasticity
should have more moderate effects.

In summary, we have shown how populations of excitatory
and inhibitory neurons can learn to efficiently represent a signal
spike by spike. This type of unsupervised learning, which includes
both principal and independent component analysis as special
cases \cite{hyvarinen2004independent}, has previously been
studied largely in rate networks
\cite{oja1982simplified,bell1995information,vertechi2014,chklovskii2015}. Implementations
that seek to mimic biology by assuming spiking neurons,
recurrent network architectures, or local learning rules
have been largely limited to heuristic or approximative approaches
\cite{savin2010independent,clopath2010connectivity,zylberberg2011sparse,king2013inhibitory}.
Using a rigorous top-down approach, we have here
derived biologically plausible rules that are guarantueed to
converge to a specific connectivity and achieve a
maximally efficient code. Importantly, single spikes are
not to be considered as random samples from a rate, but
are rather an integral part of a metabolically efficient brain.

\begin{figure}
\centering
\includegraphics[width=13cm]{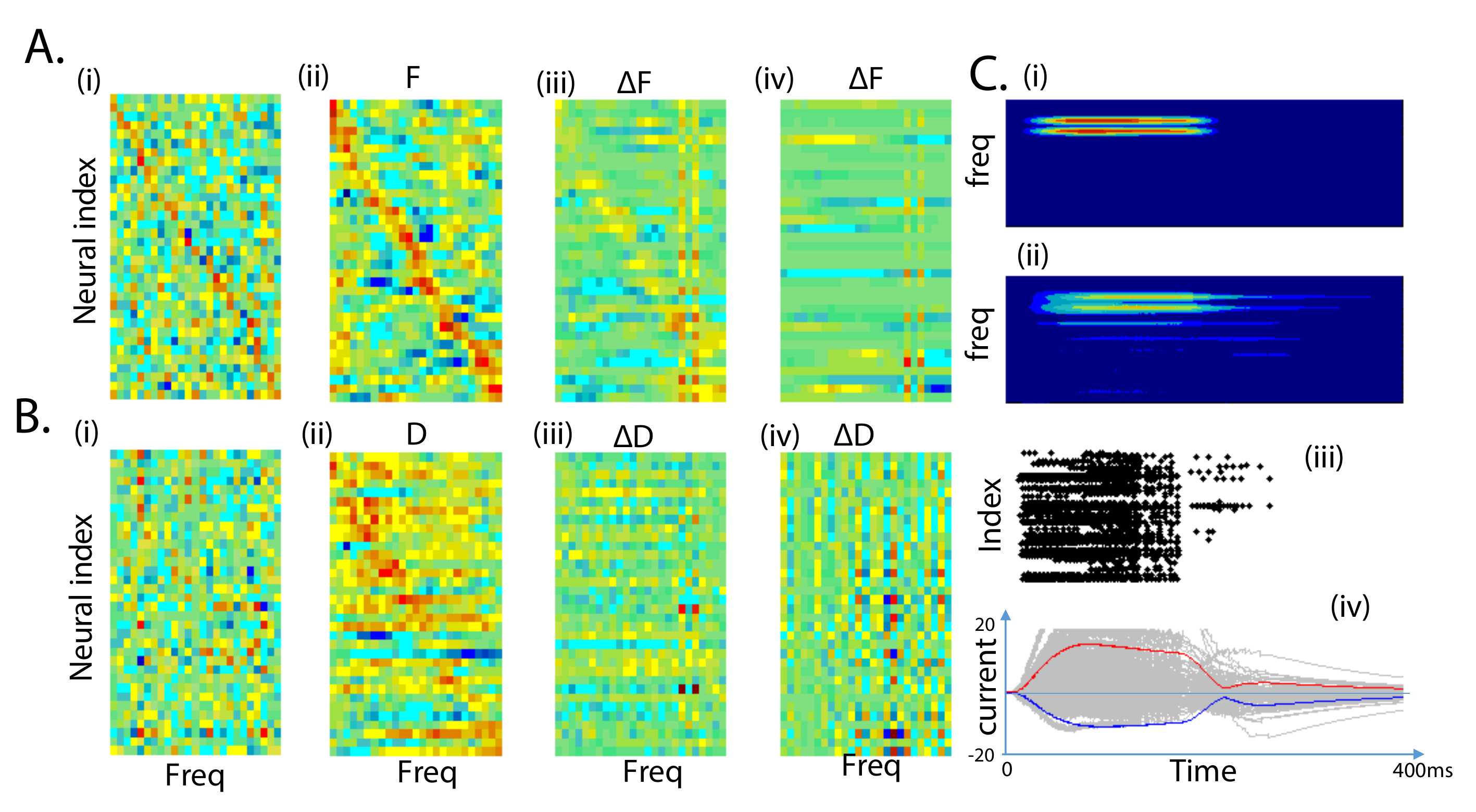}

\caption{Feedforward and recurrent connection structure before and after learning speech sound.
{\bf A.}  (i) Feed-forward weights (“STRFs”) of more neurons. The initial weights are random. The diagonal appears because these were sorted according to maximal frequency. Bluish colors correspond to negative values, reddish colors to positive values. (ii) After learning, the STRFs have an excitatory sub-field, and one or two strong inhibitory subfields. This is compatible with STRFs observed in A1 cortex~\cite{Mesgarani2014}. Note that the weights have been re-sorted according to maximal frequency. (iii) After re-training with a new stimuli (see panel C(i)) the STRFs change selectively (positively and negatively) at the position of the trained frequencies. The panel represents the difference between the weights after and before the retraining. These frequency-selective changes in STRFs shape is in line with fast plastic changes of STRFs observed following behavioral training~\cite{Yin2014}. There is also a small decrease in gain at other frequencies, due to the competition with the new stimulus. (iv) When blocking the learning of the lateral connection and only learning the feed-forward connection, the STRF change in similar fashion for the trained frequencies (without change in gain). 
{\bf B.}  Same as in A, but for the decoding weights. (i)Decoding weights before learning appear random. Note that they are sorted as the STRF in A(i) to allow comparison of feedforward and decoding weights for each neuron.(ii) After learning, the decoding weights are more structured and broader than the STRFs. This is compatible to the decoding filter of speech measured in auditory cortex~\cite{Mesgarani2014}. They have been sorted as in A(ii. (iii) After re-training to the new stimulus, a small number of decoding filter (neuron) “specialize” for the new stimulus, while the other's decoding weights change only mildly. This allows the network to minimize its firing rate response to the new stimulus, while still providing an accurate representation of it. (iv)After training only the feed-forward connections, the changes in the decoder are massive and disorganized. This reflects a severe degradation in coding performance.
{\bf C.} Response of the network after re-training with the new stimuli, but only the feedforward weights, not the lateral weights. (i) The new stimulus. (ii) The estimate of the stimulus after re-training is poorer than it was before (see fig 4 in main text). (iii) The firing rates have massively increased. (iv) The balance between excitation and inhibition have worsened. Thus, we predict that specifically blocking inhibitory plasticity during exposure to a new stimulus would actually degrade learning performance at the same time that it degrades the E/I balance. Note that to avoid a total failure of the network (whose firing rates eventually explodes without training the lateral connections), we divided the learning rate of the feed-forward connections by 4.}
\end{figure}

While limited here to the representation of time-varying signals,
our framework provide a solid starting point to move to other
types of computations. Indeed, a precise and efficient neural
code is a necessary condition for any precise neural computation.
For example, we showed previously that a second set of slower
connections can implement arbitrary linear dynamics in designed,
optimal networks [10]. The framework presented here can provide
crucial intuitions for the learning of these connections as well,
since it shows how to represent global errors in local quantities
such as voltages.

Apart from these theoretical advances, many
of the critical features that are hallmarks of cortical dynamics
follow naturally from our framework, even though they were
not included in the original objective.
We list four of the most important features. First, the predicted spike
trains are highly irregular and variable, which has indeed been widely reported in cortical neurons
\cite{tolhurst1983statistical,wohrer2013population}.
However, this variability is a signature of the network's
coding efficiency, rather than detrimental \cite{shadlen1998}
or purposeful noise  \cite{fiser2010,buesing2011}.
Second, despite this spike train variability, the membrane potentials
of similarly tuned neurons are strongly correlated (due
to shared inputs), as has indeed been found in various sensory areas
\cite{poulet2008internal,yu2010membrane}. Third, local
and recurrent inhibition in our network serves to balance the excitatory
feedforward inputs on a very fast time scale. Such EI balance,
in which inhibitory currents track excitatory currents on
a millisecond time scale has been found in various systems
and under various conditions \cite{isaacson2011inhibition,xue2014equalizing}.
Fourth, we have derived learning rules whose polarity
depends on the relative timing of pre-and post-synaptic spikes
(see insets in Fig.~3A). In fact, the respective sign switches
simply reflect the immediate sign reversal of the
coding error (and thus of the membrane
potential) after each new spike. As a result, most connections
display some features of the classic STDP rules, e.g. LTP
for pre-post pairing, and LTD for post-pre pairing \cite{caporale2008spike,feldman2012spike}.
The only exception are E-E connections that exhibit ``reverse
STDP'', i.e. potentiation for post-pre pairing (Fig.~3A).
Despite their simplicity, these rules are not only
spike-time dependent but also weight and voltage-dependent,
as observed experimentally \cite{clopath2010connectivity}.

Our framework thereby bridges from the essential biophysical
quantities, such as the membrane voltages of the neurons, to the
resulting population code, while providing crucial new insights
on learning and coding in spiking neural networks.

\bibliography{main}
\bibliographystyle{Science}

\newpage
\vspace*{-2cm}
\begin{center}
\includegraphics[width=15cm]{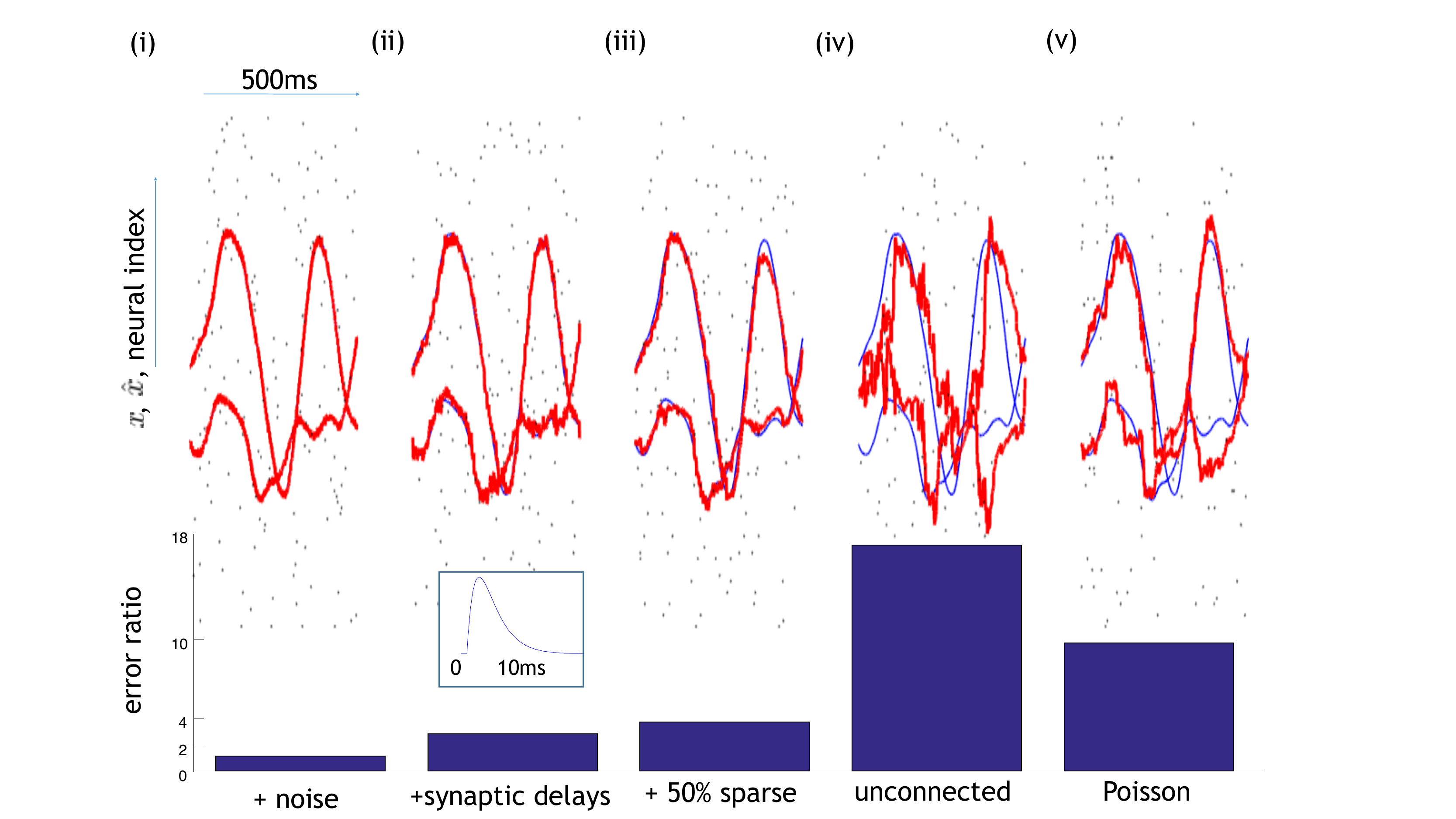}
\end{center}

\noindent
{\bf Supplementary figure S1:}
Robustness of the framework to noise, synaptic delays, and
  missing connections.  Spike trains (black dots) of 60 neurons (out
  of a total of 200), input signals (thin blue lines), network
  estimates (thick red lines), and error ratio between the perturbed
and optimal networks after learning. (i) White noise current is 
injected into each neuron. (ii) In addition to the injected noise, 
synaptic input currents are modeled with a realistic post-synaptic
potential, including a transmission delay (inset). (iii) In addition to noise and realistic synapses, 
50\% of the recurrent connections are randomly removed. 
(iv) All recurrent connections are removed so that the network is now 
composed of unconnected leaky integrate and fire neurons. (v) 
A population of independent Poisson-firing neurons with 
instantaneous firing rates identical to the network in (i),
but without learning.

\newpage

\end{document}